\documentclass[journal]{IEEEtran}

\usepackage{cite}
\usepackage{url}
\usepackage{graphicx}
\usepackage{color}
\usepackage{bm}
\usepackage{cite}
\usepackage{amsmath,amssymb,amsfonts}
\usepackage{algorithmic}
\usepackage{graphicx}
\usepackage{textcomp}
\usepackage{algorithmic}
\usepackage{algorithm}
\usepackage{array}
\usepackage[caption=false,font=normalsize,labelfont=sf,textfont=sf]{subfig}
\usepackage{stfloats}
\usepackage{url}
\usepackage{verbatim}
\usepackage{enumerate}
\usepackage{tikz}
\usepackage{bm}
\usepackage{threeparttable,booktabs}
\usepackage{blkarray}
\usepackage{amsmath}
\usepackage{multirow}

\usepackage{amsmath}
\usepackage{amsthm,amsfonts}
\usepackage{pgfplots}

\usepackage{tikz}
\usepackage[customcolors]{hf-tikz}
\usepgfplotslibrary{colorbrewer}
\usetikzlibrary{%
  3d,
  arrows,%
  shapes.misc,
  shapes.arrows,%
  chains,%
  matrix,%
	fit,%
  positioning,
  scopes,%
  decorations.pathmorphing,
  shadows,%
  decorations.markings,
  shapes.geometric,
  arrows.meta,bending,
	patterns,
	intersections, 
	pgfplots.fillbetween
}

\usepackage{tikz-3dplot}
\usepackage[pdfusetitle]{hyperref}
\hypersetup{
    colorlinks,
    linkcolor={red!50!black}, 
    citecolor={blue!50!black},
    urlcolor={blue!50!black}
}

\usepackage{cellspace}
\providecommand*{\mat}[1]{\mathbf#1}

\providecommand*{\mrm}[1]{\mathrm{#1}}

\renewcommand{\vec}[1]{{\boldsymbol#1}}

\providecommand*{\UV}[1]{\hat{\boldsymbol#1}}

\DeclareMathAccent{\ring}{\mathalpha}{operators}{"17}

\newcommand{\ie}{\textit{i.e.}\/, }

\colorlet{dpurple}{blue!50!red}
\colorlet{dblue}{blue!50!black}
\colorlet{dgreen}{green!50!black}
\colorlet{dred}{red!50!black}
\colorlet{dyellow}{yellow!50!black}
\colorlet{dorange}{orange!50!black}
\definecolor{metal}{RGB}{218,165,32}
\definecolor{diel}{RGB}{1,165,32}
\definecolor{antenna}{RGB}{100,150,162}
\definecolor{breg}{rgb}{0.2,0.6,0.8}%
\definecolor{preg}{rgb}{0.8,0.2,0.2}%
\definecolor{reg}{RGB}{218,165,32}

\graphicspath{{figures/}{tikz/}}

\makeatother

\begin{document}
\title{Fast and Rigorous Modeling of Antenna--Medium Interactions Above Planar Stratified Media via the Generalized Scattering Matrix}

\author{Chenbo Shi, Shichen Liang, Xin Gu and Jin Pan
\thanks{Manuscript received Apr. 16, 2025; revised Oct. 11, 2025. (\textit{Corresponding author: Jin Pan})}
\thanks{Chenbo Shi, Shichen Liang, Xin Gu and Jin Pan are with the School of Electronic Science and Engineering, University of Electronic Science and Technology of China, Chengdu 611731 China  (e-mail: chenbo\_shi@163.com; lscstu001@163.com; xin\_gu04@163.com; panjin@uestc.edu.cn).}
}


\maketitle

\begin{abstract}

A rigorous and computationally efficient method is presented for evaluating the reflection coefficients of antennas operating above planar layered media. The approach reformulates the problem within the framework of the antenna's generalized scattering matrix (GSM), expressed in terms of spherical vector wave functions (SVWFs). The mutual interaction between the antenna and the layered structure is modeled through spherical-to-planar vector wave transformations that incorporate the exact Fresnel reflection response of the medium, without introducing any simplifying approximations. This formulation dramatically reduces algebraic complexity and enables fast, stable numerical implementation. Excluding the one-time preprocessing required to obtain the antenna's free-space GSM, each evaluation for a given layered configuration can be completed within milliseconds---achieving several orders of magnitude speed improvement over full-wave solvers such as FEKO, while maintaining virtually identical accuracy. The proposed framework thus provides a powerful foundation for real-time electromagnetic characterization and inverse modeling involving planar layered environments.

\end{abstract}


\section{Introduction}

\IEEEPARstart{P}{arameter} measurement of planar layered media remains a topic of enduring significance, underpinning diverse applications such as pavement thickness estimation \cite{ref_thickness1,ref_thickness2,ref_thickness3}, moisture detection \cite{ref_moisture1,ref_moisture2,ref_moisture3}, density evaluation \cite{ref_density1,ref_density2}, and underground coal mining \cite{ref_coal}, among many others. These applications commonly involve solving inverse problems—that is, inferring the medium's electromagnetic parameters from the reflection coefficients measured at an antenna interface. Modern inverse techniques typically employ optimization algorithms that iteratively solve the forward problem to approximate the inverse mapping. Consequently, the computational efficiency of the forward solver becomes a decisive factor in the overall performance of such approaches.

Forward modeling of antenna behavior near planar layered media is conventionally based on the spatial-domain dyadic Green's function for layered media (DGLM), formulated via Sommerfeld integrals \cite{ref_DGLM1,ref_DGLM2,ref_DGLM3,ref_DGLM4,ref_CEM1}. This formulation underlies many commercial electromagnetic solvers, such as Altair FEKO \cite{ref_FEKO}, which employ surface integral equations over discretized antenna geometries to accurately compute reflection coefficients. However, the high computational expense of each full-wave simulation severely limits its practicality in optimization-driven workflows, which demand a large number of forward evaluations.

To mitigate this limitation, Lambot et al. introduced an elegant simplification by modeling the antenna as an equivalent point source \cite{ref_Lambot1}, using its far-field radiation pattern to approximate the reflection response from the layered medium—still within the spatial-domain DGLM framework. Although evaluating spatial-domain DGLM remains challenging due to the infinite-range Sommerfeld integrals, the point-source approximation drastically reduces computational burden by requiring evaluation at only a single spatial location. A principal drawback, however, is the far-field assumption for the antenna-medium separation. This constraint was relaxed in \cite{ref_Lambot2}, where a multi-point source model was proposed to extend accuracy into the near-field regime. While this refinement improves fidelity, it also increases the number of DGLM evaluations and introduces an additional optimization stage to determine the source positions, thereby adding nontrivial computational overhead.

More recently, Lambot and collaborators proposed a plane-wave spectrum (PWS) approach to estimate antenna reflection coefficients \cite{ref_Lambot3}. Yet, this method neglects evanescent components and multiple reflections between the antenna and the medium, restricting its applicability to scenarios involving weakly scattering antennas.

Motivated by these limitations, this paper presents a novel framework for modeling the electromagnetic interaction between antennas and planar layered media, designed specifically for frequency-domain applications that require repeated reflection-coefficient evaluations. Such scenarios encompass most ground-penetrating radar (GPR) operations and electromagnetic material characterization techniques based solely on reflection data \cite{ref_ShichenLiang,ref_XinGu}. Our method reformulates antenna transmission, reception, and scattering within the generalized scattering matrix (GSM) framework, expressed in terms of discrete spherical vector wave functions (SVWFs) rather than continuous plane-wave spectra. The antenna-medium interaction is modeled algebraically through transformations between SVWFs and planar vector wave functions (PVWFs) \cite{ref_SWF2PWF1,ref_SWF2PWF2,ref_SWF2PWF3}.

Although this SVWF-PVWF-SVWF transformation may appear mathematically intricate, the computational effort remains remarkably modest. The transformation matrix involves Sommerfeld integrals, but only over very finite integration limits---reducing the complexity to that of standard definite integrals. Compared with the double-integral operations in the spectral-domain method of \cite{ref_Lambot3}, our formulation requires only single integrals. Moreover, the inherent symmetry and sparsity of the transformation matrix yield substantial numerical acceleration.

As further demonstrated in this work, the SVWF-based formulation offers a compact representation that enables the inclusion of multiple reflections at minimal computational cost. This results in high-fidelity modeling without compromising efficiency. Excluding the one-time preprocessing cost of free-space antenna characterization, the proposed method achieves speedups of several orders of magnitude over full-wave simulations in FEKO while maintaining near-identical accuracy. This dramatic improvement in forward-model efficiency holds significant promise for accelerating modern inverse approaches involving planar layered media. 

\section{Spherical Vector Waves, Planar Vector Waves and Transformations}
\label{SecII}

A core foundation of this paper is the use of PVWFs and SVWFs, which are extensively utilized in \cite{ref_SWF2PWF1,ref_SWF2PWF2,ref_SWF2PWF3, ref_Kristensson_booklet}. We adopt the time convention \( e^{\mathrm{j}\omega t} \) and define the PVWF \( \boldsymbol{\phi }_i\left( \alpha ,\beta ,\boldsymbol{r} \right) =\boldsymbol{\phi }_i\left( \UV{\gamma},\boldsymbol{r} \right)\) as follows:
\begin{equation}
  \begin{split}
    &\boldsymbol{\phi }_1\left( \UV{\gamma},\boldsymbol{r} \right) =\frac{\mathrm{j}}{4\pi}\UV{\beta}e^{-\mathrm{j}k\UV{\gamma}\cdot \boldsymbol{r}}
    \\
    &\boldsymbol{\phi }_2\left( \UV{\gamma},\boldsymbol{r} \right) =-\frac{1}{4\pi}\UV{\alpha}e^{-\mathrm{j}k\UV{\gamma}\cdot \boldsymbol{r}}.
\end{split}
\end{equation}
Here, \(i=1\) represents the TE wave, and \(i=2\) represents the TM wave. The unit vectors \( \UV{\alpha}, \UV{\beta}, \UV{\gamma} \) are related to the elevation angle \( \alpha \) and azimuth angle \( \beta \) as follows:
\begin{equation}
  \begin{split}
    &\UV{\alpha}=\UV{x}\cos \alpha \cos \beta +\UV{y}\cos \alpha \sin \beta -\UV{z}\sin \alpha 
    \\
    &\UV{\beta}=-\UV{x}\sin \beta +\UV{y}\cos \beta 
    \\
    &\UV{\gamma}=\UV{x}\sin \alpha \cos \beta +\UV{y}\sin \alpha \sin \beta +\UV{z}\cos \alpha .
  \end{split}
\end{equation}

\begin{figure}[!t] 
  \centering
  \subfloat[]{\includegraphics[]{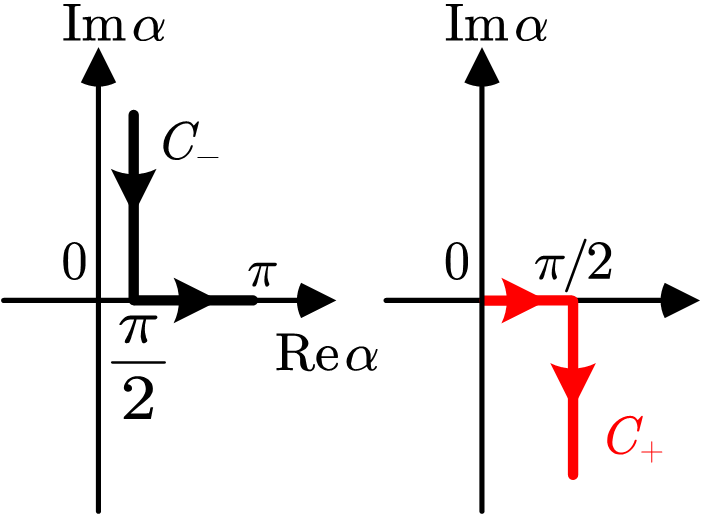}}
  \hfil
  \subfloat[]{\includegraphics[]{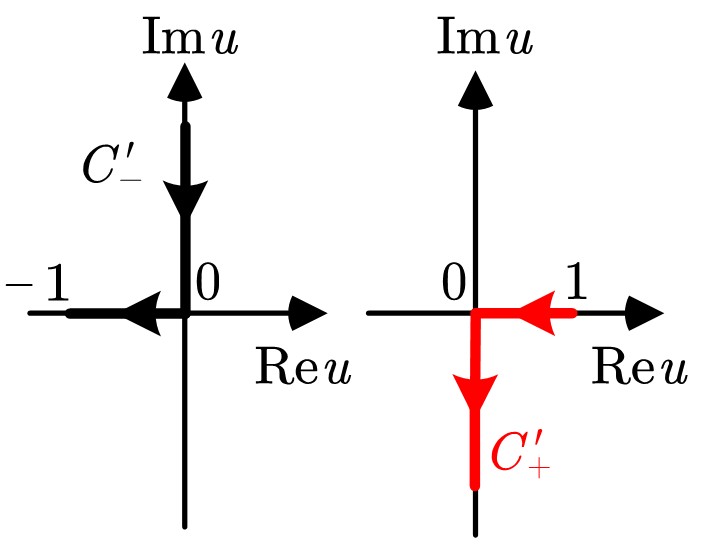}}
  \caption{Complex contours. (a) Contour of angle $\alpha$. (b) Contour of $u=\cos\alpha$. }
\label{f_contours}
\end{figure}

The azimuth angle \( \beta \) ranges from \([0, 2\pi]\), while the elevation angle \( \alpha \) spans \( C_+ \cup C_- \). The contours \( C_+ \) and \( C_- \) are illustrated in Fig. \ref{f_contours}a. For propagating waves, \( \alpha \) lies along the real axis between 0 and \( \pi \).

In this paper, we also employ SVWFs \( \boldsymbol{u}_{n}^{(p)}(\boldsymbol{r}) \), derived from real-valued spherical harmonics \cite[Appendix A]{ref_hybrid}. Here, \( p=1,4 \) denote regular and outgoing waves, respectively. The subscript \( n \rightarrow \tau \sigma ml \) indicates polarization type $\tau={1,2}$ (TE, TM), parity $\sigma={e,o}$ (even, odd), angular degree $l={1,2,\dots,L_{\max}}$, and order $m={0,1,\dots,l}$. 

Both SVWFs and PVWFs satisfy the homogeneous vector wave equation, and therefore, they can be converted into one another \cite{ref_Kristensson_booklet}:
\begin{equation}
  \label{eq3}
  \begin{split}
   &\boldsymbol{u}_{n}^{\left( 1 \right)}\left( \boldsymbol{r} \right) =\sum_i{\int_0^{\pi}{\mathrm{d}\UV{\gamma}B_{ni}\left( \UV{\gamma} \right) \boldsymbol{\phi }_i\left( \UV{\gamma},\boldsymbol{r} \right)}}
  \\
  &\boldsymbol{\phi }_i\left( \UV{\gamma},\boldsymbol{r} \right) =\sum_n{B_{ni}^{\dagger}\left( \UV{\gamma} \right) \boldsymbol{u}_{n}^{\left( 1 \right)}\left( \boldsymbol{r} \right)}
  \\
  &\boldsymbol{u}_{n}^{\left( 4 \right)}\left( \boldsymbol{r} \right) =2\sum_i{\int_{C_{\pm}}{\mathrm{d}\UV{\gamma}B_{ni}\left( \UV{\gamma} \right) \boldsymbol{\phi }_i\left( \UV{\gamma},\boldsymbol{r} \right)}},z\gtrless 0.
  \end{split}
\end{equation}
Here, \( z>0 \) corresponds to \( C_+ \), and \( z<0 \) corresponds to \( C_- \). The integral notation is defined as \(\int_C{\mathrm{d}\UV{\gamma}}=\int_C{\sin \alpha \mathrm{d}\alpha \int_0^{2\pi}{\mathrm{d}\beta}}\). The function \( B_{ni}(\UV{\gamma}) \) is given by:
\begin{equation}
  B_{ni}\left( \UV{\gamma} \right)=B_{ni}\left( \alpha,\beta \right) =B_{ni}\left( u \right) A_{ni}\left( \beta \right) ,\quad u=\cos \alpha 
\end{equation}
where
\begin{equation*}
  \begin{split}
  &B_{ni}\left( u \right) =\mathrm{j}^l\left[ \delta _{\tau i}\mathrm{j}\Delta _{l}^{m}\left( u \right) -\delta _{\tau \bar{i}}\pi _{l}^{m}\left( u \right) \right] 
  \\
  &A_{ni}\left( \beta \right) =\delta _{\tau i}\begin{cases}
	\cos \left( m\beta \right)\\
	\sin \left( m\beta \right)\\
  \end{cases}+\delta _{\tau \bar{i}}\begin{cases}
	-\sin \left( m\beta \right)\\
	\cos \left( m\beta \right)\\
  \end{cases}.
  \end{split}
\end{equation*}
Here, the upper (lower) form of $A_{ni}(\beta)$ applies for even (odd) parity $\sigma=e(o)$. The \( \delta \) is the Kronecker delta. A bar over $i$ swaps the TE/TM types of PVWFs (\textit{i.e.}, \( \bar{i} = 3 - i \)). \( B_{ni}^{\dagger}(\UV{\gamma}) \) represents the same expression as \( B_{ni}(\UV{\gamma}) \) but with all explicit \( \mathrm{j} \) replaced by \( -\mathrm{j} \). The auxiliary functions
\begin{equation}
  \begin{split}
    &\Delta _{l}^{m}\left( u \right) =-\frac{\sqrt{1-u^2}}{\sqrt{l\left( l+1 \right)}}\frac{\mathrm{d}}{\mathrm{d}u}\tilde{P}_{l}^{m}\left( u \right) \\
    &\pi _{l}^{m}\left( u \right) =-\frac{m}{\sqrt{l\left( l+1 \right)}\sqrt{1-u^2}}\tilde{P}_{l}^{m}\left( u \right)
  \end{split}
\end{equation}
where $\tilde{P}_l^m(u)$ represents the normalized associated Legendre functions \cite{ref_legendre_code}. An efficient algorithm for evaluating these functions for all $l$ less than $L_{\max}$ is provided in \cite{ref_legendre_code}. This can also be extended to evaluate $\tilde{P}_l^m(u)$ for imaginary $u$.

A key integral identity used later this work is:
\begin{equation}
  \label{equ6}
  \int_0^{2\pi}{A_{ni}\left( \beta \right) A_{n^{\prime}i}\left( \beta \right) \mathrm{d}\beta}=\delta _{mm^{\prime}}I_{\tau \sigma ,\tau ^{\prime}\sigma ^{\prime}}^{i}
\end{equation}
where \(I_{\tau \sigma ,\tau ^{\prime}\sigma ^{\prime}}^{i}\) is nonzero only in the following cases:
\begin{equation*}
  \begin{split}
  &I_{\tau \sigma ,\tau \sigma}^{i}=\pi \left[ 1+\left( -1 \right) ^{i+\tau +\sigma}\delta _{m0} \right] 
  \\
  &I_{\tau \sigma ,\tau ^{\prime}\sigma ^{\prime}}^{i}=\pi \left[ \left( -1 \right) ^{i+\tau +\sigma}+\delta _{m0} \right] ,\tau \ne \tau^ \prime\land \sigma \ne \sigma ^\prime
  \end{split}
\end{equation*}

\section{Interactions between Antenna and Planar Stratified Media}

\subsection{Expansion of the Fields into Spherical Vector Waves}

\begin{figure}[!t]
  \centering
  \includegraphics[]{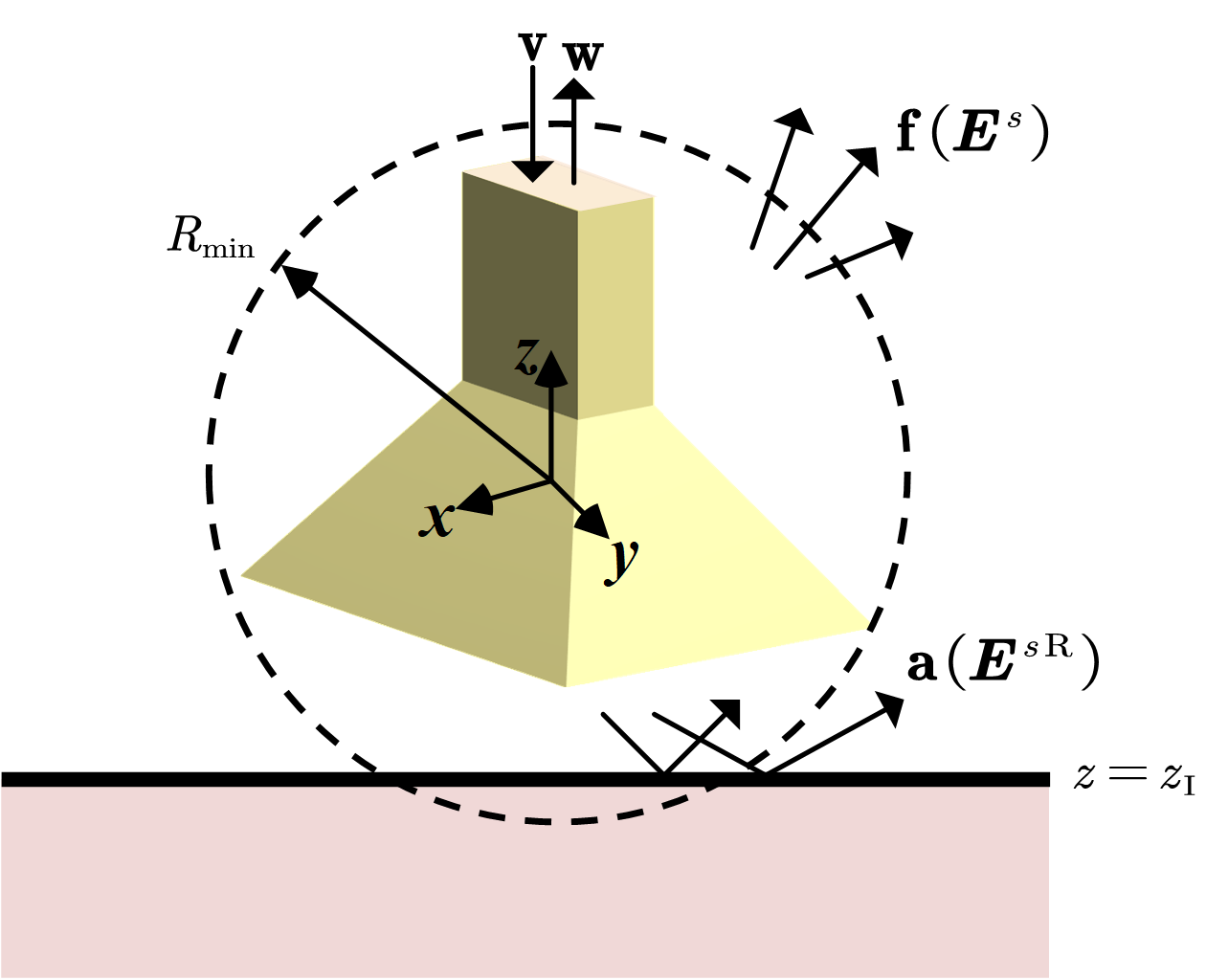}
  \caption{An antenna radiating above a planar interface. The origin of the reference coordinate system is set at the center of the antenna. Vectors $\mathbf{v}$ and $\mathbf{w}$ denote the expansion coefficients of the inward and outward waveguide eigenmodes of the antenna, respectively. $\mathbf{f}$ represents the coefficients of the outgoing SVWFs, and $\mathbf{a}$ those of the regular SVWFs. In this work, $\mathbf{f}$ is associated with the antenna-generated field $\vec{E}^s$, while $\mathbf{a}$ corresponds to the reflected field $\vec{E}^{sR}$.}
  \label{f_Antenna_Interface}
\end{figure}

Consider an antenna radiating above a planar interface (\(z = z_{\mathrm{I}}\), $z_{\mathrm{I}} < 0$), as illustrated in Fig.~\ref{f_Antenna_Interface}. Assuming that the antenna is actively excited solely by the eigenmodes $\mathbf{v}$ within its feeding waveguide, the total field in the half-space above the interface can be expressed as
\begin{equation}
  \vec E=\vec E^s+\vec E^{sR}
\end{equation}
where $\boldsymbol{E}^s$ is the field generated by the antenna, and $\vec{E}^{sR}$ represents its reflection from the planar interface.

In the antenna-centered coordinate system, $\vec{E}^s$ can be conveniently expanded as a superposition of outgoing SVWFs:
\begin{equation}
  \label{eqn8}
  \boldsymbol{E}^s\left( \boldsymbol{r} \right) =\sum_n{f_n\boldsymbol{u}_{n}^{\left( 4 \right)}\left( \boldsymbol{r} \right)}.
\end{equation}
However, obtaining an equivalent SVWF expansion for the reflected field $\vec{E}^{sR}$ is less straightforward, as it requires a transformation between SVWFs and PVWFs.To this end, consider each component $\boldsymbol{u}_{n}^{(4)}\left( \boldsymbol{r} \right)$ in \eqref{eqn8}, which can be expressed in terms of PVWFs through \eqref{eq3} (for $z < 0$) as
\begin{equation}
  \label{eqn9}
  \boldsymbol{u}_{n}^{\left( 4 \right)}\left( \boldsymbol{r} \right) =2\sum_i{\int_{C_-}{\mathrm{d}\UV{\gamma}B_{ni}\left( \UV{\gamma} \right) \boldsymbol{\phi }_i\left( \alpha ,\beta ,\boldsymbol{r} \right)}}.
\end{equation}

The reflected counterpart of this wave, $\boldsymbol{u}_{n}^{\mathrm{R}}\left( \boldsymbol{r} \right)$, can be obtained by substituting each PVWF \(\boldsymbol{\phi }_i(\alpha ,\beta ,\boldsymbol{r})\) with its reflected form $\boldsymbol{\phi }^{\mathrm{R}}_i$:
\begin{equation}
  \boldsymbol{\phi }^{\mathrm{R}}_i=\rho _i^\text{Ref}\left( \alpha \right) \boldsymbol{\phi }_i\left( \pi -\alpha ,\beta ,\boldsymbol{r} \right) 
\end{equation}
where $\rho _i^\text{Ref}\left( \alpha \right)=\rho _i^1(\alpha)e^{-2\mathrm{j}k\UV{\gamma}\cdot \boldsymbol{d}}$, and $\rho _i^1(\alpha)$ is the well-known Fresnel reflection coefficient (see Appendix \ref{app_A}). The phase term \(e^{-2\mathrm{j}k\UV{\gamma}\cdot \boldsymbol{d}}\) accounts for the path difference referenced to the antenna origin, with \(\UV{\gamma}\cdot \boldsymbol{d} = z_{\mathrm{I}}\cos \alpha\).

Thus, the PVWF representation of $\boldsymbol{u}_{n}^{\mathrm{R}}\left( \boldsymbol{r} \right)$ can be written as
\begin{equation}
  \boldsymbol{u}_{n}^{\mathrm{R}}\left( \boldsymbol{r} \right) =2\sum_i{\int_{C_-}{\mathrm{d}\UV{\gamma}B_{ni}\left( \UV{\gamma} \right) \rho _i^\text{Ref}\left( \alpha \right)\boldsymbol{\phi }_i\left( \pi -\alpha ,\beta ,\boldsymbol{r} \right)}}.
\end{equation}

Because $\boldsymbol{\phi }_i\left( \pi -\alpha ,\beta ,\boldsymbol{r} \right)$ can be reconverted into SVWFs through \eqref{eq3}, one has
\begin{equation}
  \begin{split}
  \boldsymbol{\phi }_i\left( \pi -\alpha ,\beta ,\boldsymbol{r} \right) &=\sum_{n^{\prime}}{B_{n^{\prime}i}^{\dagger}\left( \pi -\alpha ,\beta \right) \boldsymbol{u}_{n^{\prime}}^{\left( 1 \right)}\left( \boldsymbol{r} \right)}
  \\
  &=\sum_{n^{\prime}}{\left( -1 \right) ^{a'}B_{n^{\prime}i}^{\dagger}\left( \UV{\gamma} \right) \boldsymbol{u}_{n^{\prime}}^{\left( 1 \right)}\left( \boldsymbol{r} \right)}
  \end{split}
\end{equation}
where $a^{\prime}=l^{\prime}+m^{\prime}+\tau ^{\prime}+i+1$. Accordingly, $\boldsymbol{u}_{n}^{\mathrm{R}}\left( \boldsymbol{r} \right) $ can be expanded as a superposition of regular SVWFs:
\begin{equation}
  \boldsymbol{u}_{n}^{\mathrm{R}}\left( \boldsymbol{r} \right) =\sum_{n^{\prime}}{\mathcal{W} _{nn^{\prime}}\boldsymbol{u}_{n^{\prime}}^{\left( 1 \right)}\left( \boldsymbol{r} \right)}
\end{equation}
with the $nn'$-th element of the transformation matrix $\mathcal{W}$ given by
\begin{equation}
  \label{eqn14}
  \mathcal{W} _{nn^{\prime}}=2\sum_i{\left( -1 \right) ^{a^{\prime}}\int_{C_-}{\mathrm{d}\UV{\gamma}B_{ni}\left( \UV{\gamma} \right) \rho _i^\text{Ref}\left( \alpha \right)B_{n^{\prime}i}^{\dagger}\left( \UV{\gamma} \right)}}.
\end{equation}

Using these definitions, the reflected field \( \boldsymbol{E}^{s\mathrm{R}}(\boldsymbol{r}) \) can be expressed as:
\begin{equation}
  \label{eqn15}
  \begin{split}
    &\boldsymbol{E}^{s\mathrm{R}}\left( \boldsymbol{r} \right) =k\sqrt{Z_0}\sum_n{f_n\boldsymbol{u}_{n}^{\mathrm{R}}\left( \boldsymbol{r} \right)}
    \\
    &=k\sqrt{Z_0}\sum_n{\sum_{n^{\prime}}{f_n\mathcal{W} _{nn^{\prime}}\boldsymbol{u}_{n^{\prime}}^{\left( 1 \right)}\left( \boldsymbol{r} \right)}}
  \end{split}
\end{equation}
or more compactly,
\begin{equation}
  \label{eqn16}
  \boldsymbol{E}^{s\mathrm{R}}\left( \boldsymbol{r} \right) =k\sqrt{Z_0}\sum_n{a_n\boldsymbol{u}_{n}^{\left( 1 \right)}\left( \boldsymbol{r} \right)}.
\end{equation}
where, by exchanging $n,n'$ in \eqref{eqn15} and comparing with \eqref{eqn16}, we obtain the matrix relation
\begin{equation}
  \label{eq21}
  \mathbf{a}=\mathcal{W} ^t\mathbf{f}.
\end{equation}

\subsection{Solution of the Fields via Antenna Generalized Scattering Matrix}

The GSM for antennas in free space---formulated in the source-scattering representation\footnote{The standard formulation expresses the electromagnetic field in terms of outgoing and ingoing waves, with the GSM consisting only of the blocks $\mathbf{\Gamma}$, $\mathbf{R}$, $\mathbf{T}$, and $\mathbf{S}$. In practical scattering problems, however, the field is more conveniently decomposed into scattered and incident components, represented by outgoing and regular spherical waves, respectively. Since a regular wave can be expressed as one half of the sum of ingoing and outgoing waves, the identity matrix is subtracted from $\mathbf{S}$, and a factor of $1/2$ appears in \eqref{eqn18}.}---relates the inward and incident vectors $\mat v,\mat a$ to the outward and scattered vectors $\mat w,\mat f$ \cite[Ch. 2.3.5]{ref_Spherical_near_field}, \cite{ref_GSM_source_scattering,ref_myGSM}:
\begin{equation}
  \label{eqn18}
  \begin{bmatrix}
    \mathbf{\Gamma }&		\frac{1}{2}\mathbf{R}\\
    \mathbf{T}&		\frac{1}{2}\left( \mathbf{S}-\mathbf{1} \right)\\
  \end{bmatrix}
  \begin{bmatrix}
    \mathbf{v}\\
    \mathbf{a}\\
  \end{bmatrix}  = \begin{bmatrix}
    \mathbf{w}\\
    \mathbf{f}\\
  \end{bmatrix}.
\end{equation}
Here, $\mathbf{\Gamma}$ denotes the antenna's port S-parameter matrix, $\mathbf{R}$ the receiving matrix, $\mathbf{T}$ the transmitting matrix, and $\mathbf{S}$ the scattering matrix. The symbol $\mathbf{1}$ represents the identity matrix.

Substituting $\mathbf{a}=\mathcal{W}^t \mathbf{f}=\mathcal{W}\mathbf{f}$ into the second row of \eqref{eqn18} yields
\begin{equation}
  \mathbf{Tv}+\frac{1}{2}\left( \mathbf{S}-\mathbf{1} \right) \mathcal{W} \mathbf{f}=\mathbf{f}
\end{equation}
from which $\mathbf{f}$ can be solved as
\begin{equation}
  \label{eq24}
  \mathbf{f}=\left[ \mathbf{1}-\tfrac{1}{2}\left( \mathbf{S}-\mathbf{1} \right) \mathcal{W} \right] ^{-1}\mathbf{Tv}.
\end{equation}
Substituting this expression into the first row of \eqref{eqn18} gives
\begin{equation}
  \left\{ \mathbf{\Gamma }+\tfrac{1}{2}\mathbf{R}\mathcal{W} \left[ \mathbf{1}-\tfrac{1}{2}\left( \mathbf{S}-\mathbf{1} \right) \mathcal{W} \right] ^{-1}\mathbf{T} \right\} \mathbf{v}=\mathbf{w}
\end{equation}
implying that the S-parameter matrix of the antenna above the layered planar medium is
\begin{equation}
  \label{eqn22}
  \mathbf{\Gamma }^c=\mathbf{\Gamma }+\tfrac{1}{2}\mathbf{R}\mathcal{W} \left[ \mathbf{1}-\tfrac{1}{2}\left( \mathbf{S}-\mathbf{1} \right) \mathcal{W} \right] ^{-1}\mathbf{T}
\end{equation}

\begin{figure}[!t]
  \centering
  \includegraphics[]{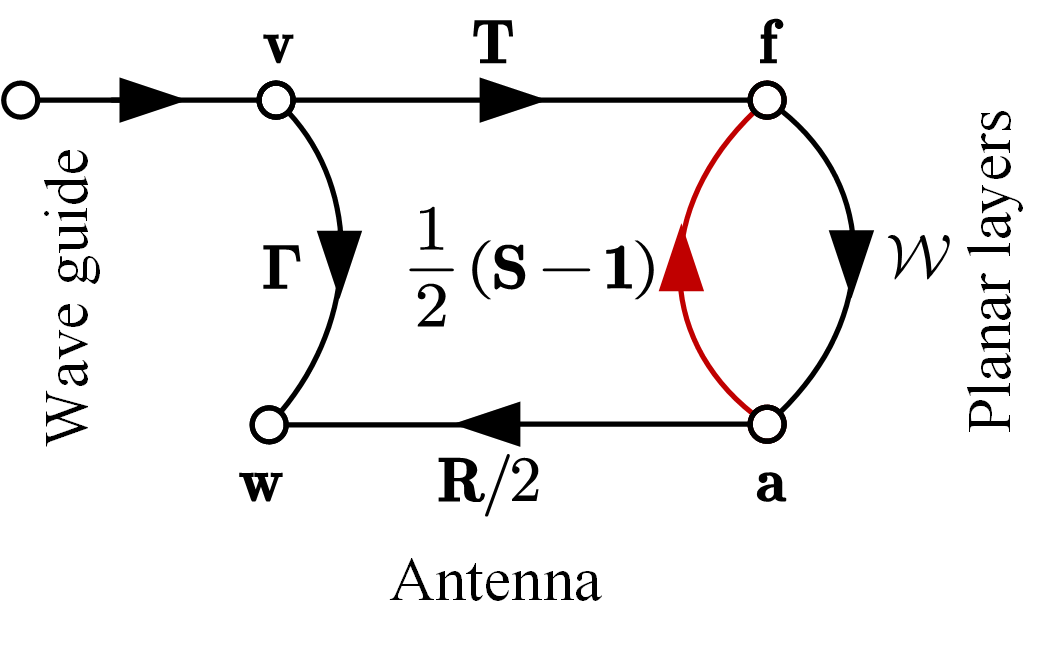}
  \caption{Signal-flow diagram illustrating the interactions between the antenna and the planar layered ground.}
  \label{f_diagram}
\end{figure}

The entire solution process can be conveniently visualized using a signal-flow graph. As illustrated in Fig.~\ref{f_diagram}, the input signal $\mathbf{v}$ propagates through the network and produces the reflected signal $\mathbf{w}$. The inverse term \( \left[ \mathbf{1} - \tfrac{1}{2}\left( \mathbf{S} - \mathbf{1} \right) \mathcal{W} \right]^{-1} \) in \eqref{eqn22} corresponds to the feedback loop on the right-hand side of the flow graph, representing the recursive interactions between antenna scattering and ground reflection. Its Neumann series expansion explicitly reveals successive orders of reflection:
\begin{equation}
  \label{eqn23}
  \begin{split}
    &\left[ \mathbf{1}-\tfrac{1}{2}\left( \mathbf{S}-\mathbf{1} \right) \mathcal{W} \right] ^{-1}\\
      &\qquad=\underbrace{\vphantom{\tfrac{1}{2}}\mathbf{1}}_{\text{1st}}+\underbrace{\tfrac{1}{2}\left( \mathbf{S}-\mathbf{1} \right) \mathcal{W}}_{\text{2nd}} +\underbrace{\left[ \tfrac{1}{2}\left( \mathbf{S}-\mathbf{1} \right) \mathcal{W} \right] ^2}_{\text{3rd}}+\cdots
  \end{split}
\end{equation}
where the first, second, and third terms correspond to the 1st-, 2nd- and 3rd-order reflections, respectively.

If multiple reflections between the antenna and the interface can be neglected, the red feedback path in Fig.~\ref{f_diagram} may be omitted, leaving only the ``1st'' term (the identity matrix $\mathbf{1}$) in \eqref{eqn23}. This simplification is valid only when antenna scattering is weak ($\mathbf{S} \to \mathbf{1}$). For enhanced accuracy, higher-order terms should be retained, or alternatively, the full inverse \( \left[ \mathbf{1} - \tfrac{1}{2}\left( \mathbf{S} - \mathbf{1} \right) \mathcal{W} \right]^{-1} \) should be evaluated directly.

A noteworthy feature of \eqref{eqn22} is that the matrices $\mathbf{\Gamma}$, $\mathbf{R}$, $\mathbf{T}$, and $\mathbf{S}$ describe the intrinsic behavior of the antenna in free space, whereas $\mathcal{W}$ depends solely on the layered medium and the interface position $z_{\mathrm{I}}$, independent of antenna geometry. Consequently, when either $z_{\mathrm{I}}$ or the medium parameters vary, only $\mathcal{W}$ must be recomputed---a process that is computationally lightweight and can typically be performed within milliseconds, as further discussed in Section~\ref{Sec_IVB}.

\section{Discussion on Computational Aspects}
\subsection{Generalized scattering matrix}
\label{SecIV_A}

The proposed framework requires precomputing the free-space antenna GSM blocks $\mathbf{\Gamma}, \mathbf{R}, \mathbf{T}, \mathbf{S}$. Following \cite[equation (45)]{ref_myGSM}, these matrices can be obtained via the method of moments (MoM):
\begin{equation}
  \label{eqn24}
  \begin{bmatrix}
	\mathbf{\Gamma }&		\mathbf{R}\\
	\mathbf{T}&		\mathbf{S}\\
\end{bmatrix} =-2\tilde{\mathbf{P}}^{\mathrm{E}}\left( \mathbf{Z}^{\mathrm{E}} \right) ^{-1}\tilde{\mathbf{P}}^{\mathrm{E}t}+\mathbf{1}
\end{equation}
Here, $\mathbf{Z}^{\mathrm{E}}$ is the impedance matrix and $\tilde{\mathbf{P}}^{\mathrm{E}}$ is the projection operator associated with the waveguide eigenmodes and SVWFs, of sizes $r\times r$ and $(e+j)\times r$, respectively. The parameter $r$ is the number of Rao--Wilton--Glisson (RWG) basis functions \cite{ref_RWG} used to discretize the antenna surface (set by the electrical size and geometric complexity), while $e$ is the number of non-cutoff feed-waveguide modes (typically small). The maximum SVWF degree $L_{\max}$ is governed by the radius $R_{\min}$ of the smallest circumscribing sphere and is independent of geometric detail \cite{ref_Sph_deg_trunction}:
\begin{equation}
  \label{eqn25}
  L_{\max}=\lceil kR_{\min}+ 2 \sqrt[3]{kR_{\min}}+3 \rceil
\end{equation}
leading to $j = 2L_{\max}(L_{\max}+2)$ SVWFs.

Consider a PEC square sheet of side length $\ell$. In this idealized case, $r$ and $j$ both scale as $(k\ell)^2$. The costs of assembling $\mathbf{Z}^{\mathrm{E}}$ and $\tilde{\mathbf{P}}^{\mathrm{E}}$ scale as $(k\ell)^4$, whereas the dominant cost in \eqref{eqn24} is inverting $\mathbf{Z}^{\mathrm{E}}$, which scales as $(k\ell)^6$.

Relative to a conventional free-space radiation analysis, the incremental effort to form the GSM arises primarily from evaluating $\tilde{\mathbf{P}}^{\mathrm{E}}$ and carrying out the matrix products in \eqref{eqn24}. Because these steps are typically far less expensive than assembling $\mathbf{Z}^{\mathrm{E}}$ and computing its inverse, the overall computational burden is essentially unchanged. In short, within a classical MoM workflow, forming the GSM generally does not add significant cost.

As a representative example, consider the multimode horn in \cite{ref_myGSM}. The surface current is discretized with 5904 RWG basis functions. With $R_{\min}\approx146$~mm and $kR_{\min}=9.785$, we obtain $L_{\max}=17$ and thus 646 SVWFs. Under these settings, computing $\tilde{\mathbf{P}}^{\mathrm{E}}$ takes $0.65$~s, assembling $\mathbf{Z}^{\mathrm{E}}$ and performing its LU factorization take $3.72$~s, and evaluating \eqref{eqn24} takes $0.26$~s, for a total of $4.63$~s per frequency point. All timings were obtained on a workstation with an AMD Ryzen 9 9950X (16 cores) and 64~GB RAM using 15-thread parallelization.

When $k\ell$ becomes very large, matrix-free acceleration techniques such as the multilevel fast multipole method (MLFMM) are typically required to handle the radiation problem efficiently. Under this framework, the cost of the matrix-vector product $\mathbf{Z}^{\mathrm{E}}\cdot\mathbf{x}$ is reduced to approximately $(k\ell)^2\log(k\ell)$, and iterative solvers can be used to evaluate \eqref{eqn24}. The overall computational complexity then scales with the number of iterations $N_{\mathrm{it}}$, the number of SVWFs $j\sim(k\ell)^2$, and the matrix-vector cost, \ie $N_{\mathrm{it}}(k\ell)^4\log(k\ell)$. In contrast, a conventional free-space radiation problem can be reduced to $N_{\mathrm{it}}(k\ell)^2\log(k\ell)$.

Therefore, for electrically very large antennas, the explicit construction and inversion of the GSM may become the dominant computational bottleneck. The current formulation, while highly efficient for small and moderately sized antennas, does not yet achieve the same level of acceleration as conventional MLFMM-based radiation analysis. Nevertheless, when suitable accelerations are available, the proposed approach retains important advantages. For example, in electrically large systems comprising multiple, well-separated antennas, the overall GSM can be assembled analytically from subcomponent GSMs \cite{ref_mySYN}, obviating the evaluation of \eqref{eqn24}.

A promising research direction is to develop matrix-free formulations that avoid the explicit computation and storage of the GSM altogether. Such approaches could preserve the physical rigor of the proposed framework while extending its scalability to electrically very large or geometrically intricate antennas. This perspective provides a natural path for future improvement and represents an exciting opportunity to broaden the applicability of the GSM-based modeling methodology.

\subsection{\texorpdfstring{Transformation matrix $\mathcal{W}$}{}}
\label{Sec_IVB}

The second major computational task is the evaluation of the transformation matrix $\mathcal{W}$. Since the integration over $\beta$ in \eqref{eqn14} can be carried out analytically (as shown in \eqref{equ6}), \eqref{eqn14} reduces to a single-integral form:
\begin{equation}
  \label{eqn26}
  \begin{split}
  \mathcal{W} _{nn^{\prime}}&=2\delta _{mm^\prime}\times\\
  &\sum_i{\left( -1 \right) ^{a^{\prime}}I_{\tau \sigma ,\tau ^{\prime}\sigma ^{\prime}}^{i}\int_{C_{-}^{\prime}}{\mathrm{d}uB_{ni}\left( u \right) \rho _{i}^{\text{Ref}}\left( u \right) B_{n^{\prime}i}^{\dagger}\left( u \right)}}
  \end{split}
\end{equation}
where $u=\cos\alpha$. The Sommerfeld integral in \eqref{eqn26} conventionally begins at \(\mathrm{j}\infty\) along the contour \(C_{-}^{\prime}\) in Fig. \ref{f_contours}b. However, recent studies have shown that this integral can be safely truncated at a very finite starting point \(\mathrm{j}\tilde{\kappa}_m \), with \( \tilde{\kappa}_m > 1 \), instead of extending to infinity. This truncation substantially reduces the computational complexity of \eqref{eqn26} while maintaining convergence throughout the region above the planar interface.

An empirical formula for determining \(\tilde{\kappa}_m\), valid for $L_{\max} \le 20$ (corresponding to antennas with diameters up to approximately $4\lambda$), is given by \cite{ref_PRA_truncation_number,ref_Rubio}:
\begin{equation}
  \label{eqn27}
  \tilde{\kappa}_m=\left( \iota L_{\max} + 1 \right) \left( kR_{\min} \right) ^{-1}+0.03kR_{\min}
\end{equation}
where the accuracy-control parameter $\iota$ is set to 0.55 in this work.

Due to the presence of the $\delta_{mm'}$ factor in \eqref{eqn26}, the number of nonzero entries in $\mathcal{W}$ increases only linearly with $kR_{\min}$, ensuring rapid evaluation. Moreover, since $\mathcal{W}$ is symmetric ($\mathcal{W} = \mathcal{W}^t$), both the computational and storage costs are effectively halved.
For instance, with $L_{\max} = 17$ and $\tilde{\kappa}_m = 1.31$, the computation of the $\mathcal{W}$ matrix using a 33-point Gauss--Legendre quadrature rule requires only about 3~ms.

\section{Numerical Examples}
\label{Sec_V}

This section presents numerical results that validate the proposed formulation across a variety of planar layered environments. The multimode horn antenna introduced in Sec.~\ref{SecIV_A} and \cite{ref_myGSM} is employed as the transmitter. The coordinate origin is located at the geometric center of the antenna, which lies 80~mm behind its radiation aperture. Over the frequency range from 3.2~GHz to 3.8~GHz, the antenna supports five propagating modes: $\text{TE}_{10}$, $\text{TE}_{20}$, $\text{TE}_{01}$, $\text{TE}_{11}$, and $\text{TM}_{11}$. The maximum degree of SVWFs is set to \( L_{\max} = 17 \), and the truncation parameter is chosen as \( \tilde{\kappa}_m = 1.31 \).

\subsection{Example 1---Perfect electric conductor half space}

\begin{figure}[!t]
  \centering
  \subfloat[]{\includegraphics[]{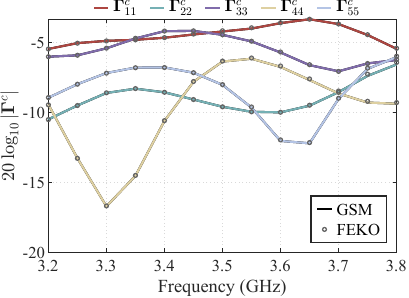}}
  \vfil
  \subfloat[]{\includegraphics[]{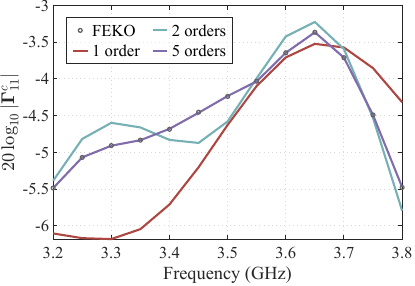}}
  \caption{S-parameters of the horn antenna above a PEC half-space environment. The interface is at $z_\mathrm{I}=-200$~mm. (a) Magnitude in dB scale. (b) $S_{11}$ computed considering different orders of reflections. } 
\label{f_S_halfspace_PEC_d200mm}
\end{figure}

The first case considers a perfect electric conductor (PEC) half-space, with the planar interface located at \( z_{\mathrm{I}} = -200 \)~mm. The distance between the horn aperture and the ground is 120~mm, corresponding to approximately \(1.28\lambda_L \), where \( \lambda_L \) represents the wavelength at 3.2~GHz. Fig. \ref{f_S_halfspace_PEC_d200mm}a shows the S-parameters for different modes. The curves labeled ``GSM'' correspond to results from the proposed method, while those labeled ``FEKO'' denote full-wave simulations from the commercial software package. Excellent agreement is observed between the two.

Figure~\ref{f_S_halfspace_PEC_d200mm}b illustrates the influence of including multiple reflection orders on numerical accuracy, using \( \mathbf{\Gamma}_{11}^{c} \) as an example. Neglecting higher-order reflections (\ie retaining only the 1st term in \eqref{eqn23}) leads to a maximum deviation of approximately 1.25~dB. Incorporating the first two reflections (the 1st and 2nd terms in \eqref{eqn23}) reduces the discrepancy to below 0.5~dB, while including up to five reflection orders yields nearly identical results to the FEKO full-wave solution. These results highlight the necessity of accounting for multiple reflections---particularly for horn antennas---since simplified approaches that ignore them may yield significant errors.

\begin{figure}[!t]
  \centering
  \includegraphics[]{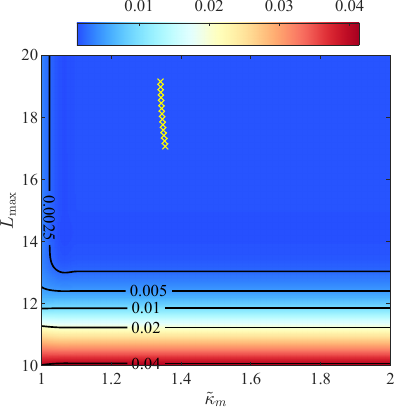}
  \caption{Maximum absolute error of $\mat\Gamma^c$ for $z_\mrm{I}=-200$~mm. The color scale (cool to warm) denotes the error magnitude (linear scale). Contours indicate equal-error regions, and yellow crosses mark $(L_{\max},\tilde{\kappa}_m)$ values determined by \eqref{eqn25} and \eqref{eqn27} at various frequencies.}
  \label{f_Err_ana_PEC_d200mm}
\end{figure}

Figure~\ref{f_Err_ana_PEC_d200mm} shows the maximum absolute error of $\mat\Gamma^c$ over the 3.2--3.8~GHz range, computed with different $L_{\max}$ and $\tilde{\kappa}_m$. A value of $L_{\max}=12$ already ensures an error below 0.01. In addition, any $\tilde{\kappa}_m>1$ provides sufficient accuracy because the horn is relatively far\footnote{The classification of ``far'' or ``near'' depends on whether the antenna-ground distance $\left| z_{\mathrm{I}} \right|$ exceeds $R_{\min}$. If $\left| z_{\mathrm{I}} \right|>R_{\min}$, the antenna is regarded as being ``far''; otherwise, it is considered ``near''.} from the interface. The empirical choices of $L_{\max}$ and $\tilde{\kappa}_m$ from \eqref{eqn25} and \eqref{eqn27}, shown as cross markers, lie within the region of very low error.

\begin{figure}[!t]
  \centering
  \includegraphics[]{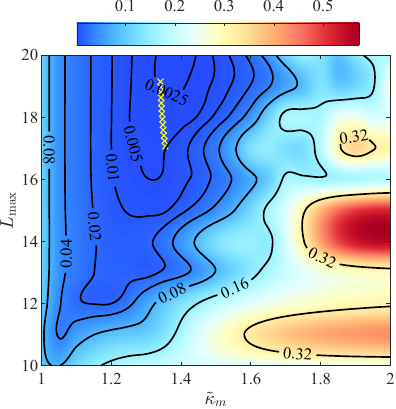}
  \caption{Maximum absolute error of $\mat\Gamma^c$ for $z_\mrm{I}=-100$~mm. The color map represents error magnitude, with contours marking equal-error regions. Yellow crosses correspond to $(L_{\max}, \tilde{\kappa}_m)$ values from \eqref{eqn25} and \eqref{eqn27}.}
  \label{f_Err_ana_PEC_d100mm}
\end{figure}

To further assess accuracy, we place the antenna close to the ground by setting \( z_{\mathrm{I}} = -100 \)~mm, corresponding to a 20~mm aperture-ground spacing (\( 0.21\lambda_L \)). Figure~\ref{f_Err_ana_PEC_d100mm} depicts the resulting error map versus $L_{\max}$ and $\tilde{\kappa}_m$. Here, the parameter selection requires slightly more care, and the empirical formulas still yield values within a low-error region. The concentration of the low-error region indicates that the results are not highly sensitive to moderate variations in either parameter, explaining the reliability of the empirical criteria.

\begin{figure}[!t]
  \centering
  \subfloat[]{\includegraphics[]{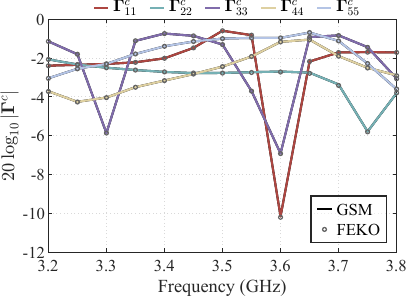}}
  \vfil
  \subfloat[]{\includegraphics[]{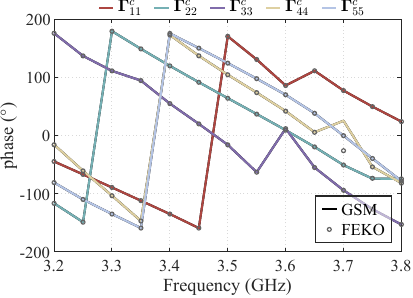}}
  \caption{S-parameters of the horn antenna in above a PEC half-space at $z_\mathrm{I}=-100$~mm. (a) Magnitude and (b) phase.}
\label{f_S_halfspace_PEC_d100mm}
\end{figure}

For instance, at $L_{\max}=17$ and $\tilde{\kappa}_m=1.31$, the magnitude and phase of $\mat{\Gamma}^c$ are shown in Figs.~\ref{f_S_halfspace_PEC_d100mm}a and \ref{f_S_halfspace_PEC_d100mm}b, respectively. The proposed method again matches the FEKO solution extremely well, confirming its accuracy and robustness even for near-ground configurations.

\subsection{Example 2---Dielectric material half space}
\label{Sec_VB}
\begin{figure}[!t]
  \centering
  \subfloat[]{\includegraphics[]{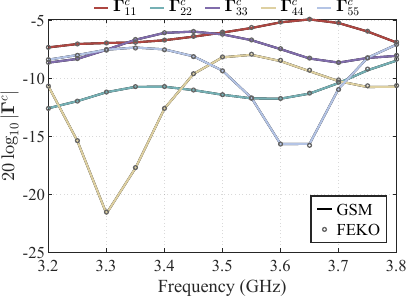}}
  \vfil
  \subfloat[]{\includegraphics[]{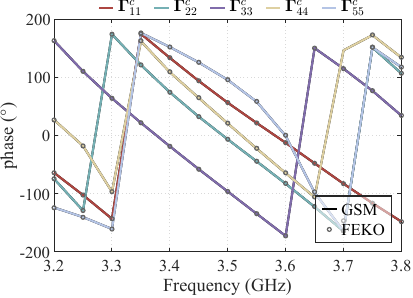}}
  \caption{S-parameters of the horn antenna above a seawater half-space ($z_\mathrm{I}=-200$~mm). (a) Magnitude. (b) Phase.} 
\label{f_Spara_halfspace_water}
\end{figure}

Next, we consider a dielectric half-space environment. The lower medium is seawater, characterized by \( \epsilon_r \approx 81 \) and \( \sigma \approx 10 \)~S/m, with the interface positioned at \( z_{\mathrm{I}} = -200 \)~mm. Figs.~\ref{f_Spara_halfspace_water}a and \ref{f_Spara_halfspace_water}b present the computed S-parameters. The proposed results are in excellent agreement with the FEKO full-wave simulations.

\begin{figure}[!t]
  \centering
  \subfloat[]{\includegraphics[]{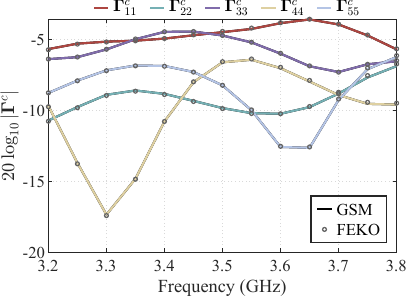}}
  \vfil
  \subfloat[]{\includegraphics[]{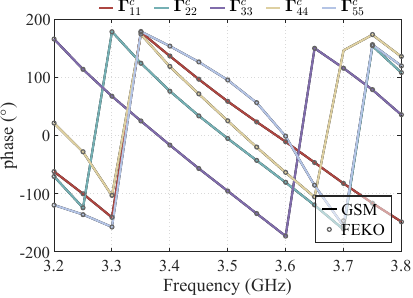}}
  \caption{S-parameters of the horn antenna in a dielectric half-space environment with $\epsilon_r=81,\sigma=500$~S/m. The interface is at $z_\mathrm{I}=-200$~mm. (a) Magnitude and (b) phase.} 
\label{f_Spara_halfspace_water_highsigma}
\end{figure}

To further examine numerical stability, a high-loss case is tested by increasing the conductivity to \( \sigma = 500 \)~S/m. The corresponding results, shown in Figs.~\ref{f_Spara_halfspace_water_highsigma}a and \ref{f_Spara_halfspace_water_highsigma}b, maintain excellent agreement with FEKO, confirming that the proposed method remains robust under high-loss conditions.

\subsection{Example 3---Layered planar medium environment}

\begin{figure}[!t]
  \centering
  \includegraphics[]{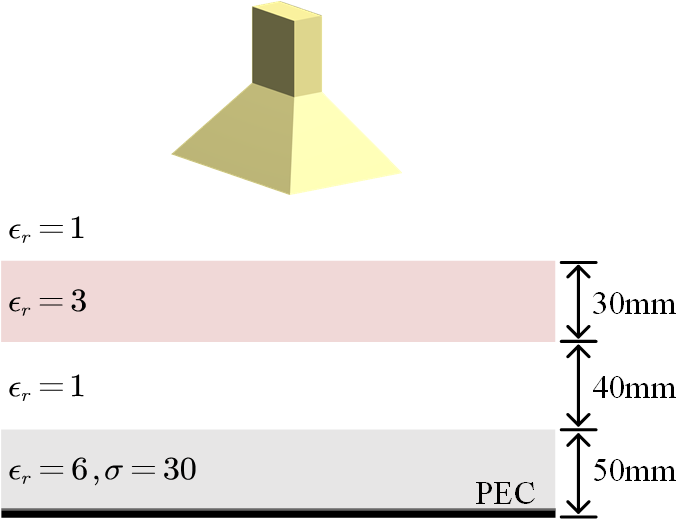}
  \caption{The planar layered model composed of five homogeneous layers.}
  \label{f_MultiLayerInfo}
\end{figure}

Finally, we examine a general multilayer planar configuration, as illustrated in Fig.~\ref{f_MultiLayerInfo}. The FEKO multilayer solver, which evaluates Sommerfeld integrals, supports only lumped-port simulations; thus, the full-wave reference is performed in free space.

The five-layer structure has lateral dimensions of \( 10\lambda_L \), ensuring quasi-infinite extent. The multilevel fast multipole method (MLFMM) is enabled in FEKO for acceleration. To facilitate MLFMM convergence, the layer conductivities are kept small, and moderate permittivities and thicknesses are chosen to reduce the mesh size (approximately \( 0.16\lambda_L \)), allowing the simulation to complete within reasonable time frame. These simplifications do not affect the generality of the conclusions. The top interface is located at \( z_{\mathrm{I}} = -200 \)~mm.

\begin{figure}[!t]
  \centering
  \subfloat[]{\includegraphics[]{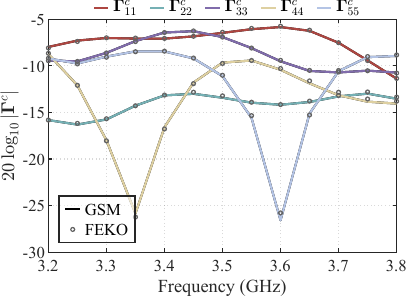}}
  \vfil
  \subfloat[]{\includegraphics[]{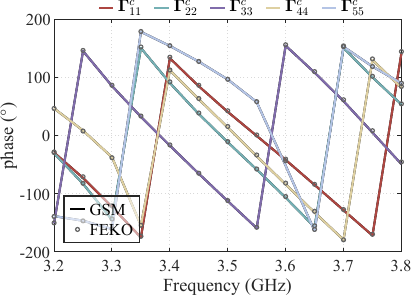}}
  \caption{S-parameters of the horn antenna in a large planar multilayer environment. The interface is at $z_\mathrm{I}=-200$~mm. (a) Magnitude and (b) phase.} 
\label{f_Spara_planar_multilayer}
\end{figure}

Figures~\ref{f_Spara_planar_multilayer}a and \ref{f_Spara_planar_multilayer}b compare the proposed results with those from the FEKO MLFMM solver. Both magnitude and phase of the antenna S-parameters show excellent agreement. Minor discrepancies at certain frequencies arise from the coarse FEKO meshing and inherent differences in numerical formulation. Overall, the results confirm that the proposed method accurately models antenna-medium interactions in complex layered environments.

\subsection{Example 4---Demonstration of the implementing time}
\label{Sec_IVD}

The total computational cost of the proposed approach comprises two primary components: (1) the precomputation of the free-space GSM, and
(2) the postprocessing that accounts for the effects of the planar layered medium.

In the configuration analyzed in the previous subsections, the precomputation of the GSM requires approximately 4.63~s (see Sec.~\ref{SecIV_A} for details). The postprocessing for the planar medium requires an additional 8.3~ms, consisting of about 3~ms for computing the transformation matrix $\mathcal{W}$---which is nearly independent of the specific layer configuration---and 5.3~ms for evaluating \eqref{eqn22}.

For reference, in the dielectric half-space scenario of Sec.~\ref{Sec_VB}, a single-frequency simulation using the \textit{Reflection Coefficient Approximation} solver in FEKO requires roughly 10.8~s. Thus, for a single evaluation, the computational cost of the proposed method is of the same order of magnitude as that of FEKO. However, when the material parameters of the medium change, FEKO must recompute the entire solution since the DGLM is altered, whereas in the proposed method only the matrix terms in the postprocessing stage need updating. Consequently, the proposed approach achieves several orders of magnitude higher efficiency when repeated evaluations over varying medium parameters are required.

It is also worth noting that FEKO's Sommerfeld-integral solver cannot be used with wave-port excitations and generally requires even longer runtimes than reflection coefficient approximation method. Moreover, for thin-layer configurations, convergence issues often arise. These comparisons collectively demonstrate the significant computational advantages and robustness of the proposed method over conventional full-wave techniques.

\section{Conclusion}

This paper has presented a novel and computationally efficient method for evaluating the reflection coefficients of antennas operating in the vicinity of planar layered media. The proposed approach models the antenna's transmitting, receiving, and scattering behavior through a generalized scattering matrix formulated in terms of SVWFs. By transforming SVWFs into PVWFs and incorporating the Fresnel reflection characteristics of the layered medium, the mutual interaction between the antenna and the stratified structure is rigorously captured. No simplifying approximations are introduced, ensuring accuracy equivalent to that of full-wave solvers such as FEKO.

The formulation features compact matrix dimensions and significantly reduced algebraic complexity. Excluding the one-time preprocessing step required for the free-space GSM, each evaluation for a given layered configuration can be completed within milliseconds---several orders of magnitude faster than conventional full-wave simulations. These advantages make the method particularly well suited for real-time and iterative applications, such as electromagnetic material characterization and inverse modeling of planar layered environments. Furthermore, the presented framework provides a unified foundation for efficiently integrating antenna-medium interactions into modern optimization and inversion algorithms.

\begin{appendices}
\section{Fresnel Reflection Coefficients for Homogeneous Planar Multi-Layers}
\label{app_A}
The reflection coefficient at the \(n\)-\(n+1\) layer interface is given by:
\begin{equation}
  \Gamma _{i}^{n}=\left( -1 \right) ^{i+1}\frac{Z_{i}^{n+1}-Z_{i}^{n}}{Z_{i}^{n+1}+Z_{i}^{n}}
\end{equation}
where \( Z_{i}^{n} \) is the wave impedance of the \(n\)-th layer medium. Specifically, for TE waves, \( Z_{1}^{n} = \frac{\omega \mu_n}{k_{z,n}} \), and for TM waves, \( Z_{2}^{n} = \frac{k_{z,n}}{\omega \epsilon_n} \). Here, \( \epsilon_n = \epsilon_{r,n} \epsilon_0 - \mathrm{j} \frac{\sigma_n}{\omega} \) and \( \mu_n = \mu_{r,n} \mu_0 \), where \( \epsilon_{r,n} \), \( \mu_{r,n} \), and \( \sigma_n \) are the relative permittivity, relative permeability, and conductivity of the \(n\)-th layer medium, respectively. For a PEC, $Z_i^n=0$, while for a perfect magnetic conductor (PMC), $Z_i^n\to\infty$. 

The longitudinal wavenumber
\begin{equation}
  k_{z,n}=k_{z,n}\left( \alpha \right) =\sqrt{k_{n}^{2}-k_{1}^{2}\sin ^2\alpha}
\end{equation}
where \( k_n=\omega\sqrt{\epsilon_n\mu_n}\) is the wavenumber in the $n$-th layer, and $k_1$ is the wavenumber for the top layer (typically free space where the antenna is placed). The convention for the square root is \( \sqrt{-1} \equiv -\mathrm{j} \).

With these notations, the Fresnel reflection coefficient at the \(n\)-\(n+1\) layer interface is expressed as:
\begin{equation}
  \rho _{i}^{n}\left( \alpha \right) =\frac{\Gamma _{i}^{n}+\Gamma _{i}^{n+1}e^{-2\mathrm{j}k_{z,n+1}h_{n+1}}}{1+\Gamma _{i}^{n+1}e^{-2\mathrm{j}k_{z,n+1}h_{n+1}}}
\end{equation}
where $h_n$ denotes the thickness of the $n$-th layer medium. For convenience, the following notations are equivalent:  \( \rho _{i}^{n}(\alpha) \), \( \rho _{i}^{n}(\cos \alpha) \), and \( \rho _{i}^{n}(u) \)---all representing the Fresnel reflection coefficient of a plane wave incident at the complex elevation angle \( \alpha \).

\end{appendices}


\begin{thebibliography}{00}

\bibitem{ref_thickness1}
A. Loizos and C. Plati, ``Accuracy of ground penetrating radar hornantenna technique for sensing pavement subsurface,'' \emph{IEEE Sensors J.}, vol. 7, no. 5, pp. 842-850, May 2007.
\bibitem{ref_thickness2}
H. Liu and M. Sato, ``In situ measurement of pavement thickness and dielectric permittivity by GPR using an antenna array,'' \emph{NDT \& E Int.}, vol. 64, pp. 65-71, Jun. 2014. 
\bibitem{ref_thickness3}
S. Zhao and I. L. Al-Qadi, ``Super-resolution of 3-D GPR signals to estimate thin asphalt overlay thickness using the XCMP method,'' \emph{IEEE Trans. Geosci. Remote Sens.}, vol. 57, no. 2, pp. 893-901, Feb. 2019.
\bibitem{ref_moisture1}
K. Grote, S. Hubbard, J. Harvey, and Y. Rubin, ``Evaluation of infiltration in layered pavements using surface GPR reflection techniques,'' \emph{J. Appl. Geophys.}, vol. 57, no. 2, pp. 129-153, Feb. 2005.
\bibitem{ref_moisture2}
J. Minet, P. Bogaert, M. Vanclooster, and S. Lambot, ``Validation of ground penetrating radar full-waveform inversion for field scale soil moisture mapping,'' \emph{J. Hydrol.}, vol. 424-425, pp. 112-123, Mar. 2012.
\bibitem{ref_moisture3}
F. M. Fernandes, A. Fernandes, and J. Pais, ``Assessment of the density and moisture content of asphalt mixtures of road pavements,'' \emph{Construct. Building Mater.}, vol. 154, pp. 1216-1225, Nov. 2017. 
\bibitem{ref_density1}
I. L. Al-Qadi, Z. Leng, S. Lahouar, and J. Baek, ``In-place hot-mix asphalt density estimation using ground-penetrating radar,'' \emph{Transp. Res. Rec., J. Transp. Res. Board}, vol. 2152, no. 1, pp. 19-27, Jan. 2010.
\bibitem{ref_density2}
S. Zhao, ``Development of GPR data analysis algorithms for predicting thin asphalt concrete overlay thickness and density,'' Ph.D. dissertation, Dept. Civil Environ. Eng., Univ. Illinois Urbana-Champaign, Champaign, IL, USA, 2018.
\bibitem{ref_coal}
S. B. Thomas and L. P. Roy, ``Impact of GPR antenna height in estimating coal layer thickness using spatial smoothing techniques,'' \emph{IET Sci., Meas. Technol.}, vol. 14, no. 10, pp. 906-912, Dec. 2020.
\bibitem{ref_DGLM1}
K. A. Michalski and D. Zheng, ``Electromagnetic scattering and radiation by surfaces of arbitrary shape in layered media, Part I: Theory,'' \emph{IEEE Trans. Antennas Propag.}, vol. 38, pp. 335-344, 1990.  
\bibitem{ref_DGLM2}
J. S. Zhao, W. C. Chew, C. C. Lu, E. Michielssen, and J. M. Song, ``Thin-stratified medium fast-multipole algorithm for solving microstrip structures,'' \emph{IEEE Trans. Microwave Theory Tech.}, vol. 46, no. 4, pp. 395-403, Apr. 1998.  
\bibitem{ref_DGLM3}
T. J. Cui and W. C. Chew, ``Fast evaluation of sommerfield integrals for em scattering and radiation by three-dimensional buried objects,'' \emph{IEEE Geosci. Remote Sens.}, vol. 37, no. 2, pp. 887-900, Mar. 1999.  
\bibitem{ref_DGLM4}
W. C. Chew, J. S. Zhao, and T. J. Cui, ``The layered medium Green's function—A new look,'' \emph{Microwave Opt. Tech. Lett.}, vol. 31, no. 4, pp. 252-255, 2001.
\bibitem{ref_CEM1}
W. C. Chew, M. S. Tong, and B. Hu, \emph{Integral Equation Methods for Electromagnetic and Elastic Waves}. in Synthesis Lectures on Computational Electromagnetics. Cham: Springer International Publishing, 2009. doi: 10.1007/978-3-031-01707-0.
\bibitem{ref_FEKO}
(2024). Altair FEKO. Altair. [Online]. Available: https://www.altair.com/feko/
\bibitem{ref_Lambot1}
S. Lambot, E. C. Slob, I. van den Bosch, B. Stockbroeckx, and M. Vanclooster, ``Modeling of ground-penetrating radar for accurate characterization of subsurface electric properties,'' \emph{IEEE Trans. Geosci. Remote Sens.}, vol. 42, no. 11, pp. 2555-2568, Nov. 2004.
\bibitem{ref_Lambot2}
S. Lambot and F. André, ``Full-wave modeling of near-field radar data for planar layered media reconstruction,'' \emph{IEEE Trans. Geosci. Remote Sens.}, vol. 52, no. 5, pp. 2295-2303, May 2014.
\bibitem{ref_Lambot3}
R. Roohi, A. R. Attari, M. S. Majedi, and S. Lambot, ``Prediction of Antenna Reflection Coefficient in Presence of Multilayer Media: A Fast Spectral Domain Approach,'' \emph{IEEE Trans. Geosci. Remote Sens.}, vol. 62, 2024.
\bibitem{ref_ShichenLiang}
S. Liang, C. Shi, and Y. Liu, ``A Matrix Pencil-Assisted Parameterized Electromagnetic Parameter Inversion Method for Half-Space Medium via a Single Antenna,'' \emph{IEEE Trans. Antennas Propag.}, vol. 73, no. 2, pp. 1018-1027, Feb. 2025.
\bibitem{ref_XinGu}
X. Gu, S. Liang, Y. Tang, C. Shi and J. Pan, ``Reflection Method for Shielding Effectiveness Measurement: Complex Materials Without Prior Knowledge,'' \emph{IEEE Trans. Antennas Propag.}, doi: 10.1109/TAP.2025.3558607.
\bibitem{ref_SWF2PWF1}
G. Kristensson, ``Electromagnetic scattering from buried inhomogeneities---a general three-dimensional formalism,'' \emph{J. Appl. Phys.}, vol. 51, no. 7, pp. 3486, 1980. 
\bibitem{ref_SWF2PWF2}
G. Videen, ``Light scattering from a sphere on or near a surface,'' \emph{J. Opt. Soc. Am. A}, vol. 8, no. 3, pp. 483, 1991. 
\bibitem{ref_SWF2PWF3}
D. W. Mackowski, ``Exact solution for the scattering and absorption properties of sphere clusters on a plane surface,'' \emph{J. Quant. Spectrosc. Radiat. Transf.}, vol. 109, no. 5, pp. 770-788, 2008.
\bibitem{ref_Kristensson_booklet}
A. Boström, G. Kristensson, and S. Ström, ``Transformation properties of plane, spherical and cylindrical scalar and vector wave functions,'' in \emph{Acoustic, Electromagnetic and Elastic Wave Scattering, Field Representations and Introduction to Scattering}, Elsevier, 1991, pp. 165-210.
\bibitem{ref_hybrid}
V. Losenicky, L. Jelinek, M. Capek, and M. Gustafsson, ``Method of Moments and T-Matrix Hybrid,'' \emph{IEEE Trans. Antennas Propag.}, vol. 70, no. 5, pp. 3560-3574, May 2022, doi: 10.1109/TAP.2021.3138265.
\bibitem{ref_legendre_code}
T. Limpanuparb and J. Milthorpe, ``Associated Legendre polynomials and spherical harmonics computation for chemistry applications,'' \emph{arXiv preprint arXiv:1410.1748}, 2014.
\bibitem{ref_Spherical_near_field}
J. E. Hansen, Ed., \emph{Spherical Near-Field Antenna Measurements.} London, U.K.: Peter Peregrinus Ltd., 1988.
\bibitem{ref_GSM_source_scattering}
A. D. Yaghjian, \emph{Near-field antenna measurements on a cylindrical surface: A source scattering-matrix formulation}, Vol. 696,  Department of Commerce, National Bureau of Standards, Institute for Basic Standards, Electromagnetics Division, 1977.
\bibitem{ref_myGSM}
C. Shi, J. Pan, X. Gu, S. Liang and L. Zuo, ``Generalized Scattering Matrix of Antenna: Moment Solution, Compression Storage and Application,'' \emph{IEEE Trans. Antennas Propag.}, vol. 73, no. 8, pp. 5791-5800, Aug. 2025.
\bibitem{ref_RWG}
S. Rao, D. Wilton, and A. Glisson, ``Electromagnetic scattering by surfaces of arbitrary shape,'' \emph{IEEE Trans. Antennas Propag.}, vol. 30, no. 3, pp. 409-418, May 1982, doi: 10.1109/TAP.1982.1142818.
\bibitem{ref_Sph_deg_trunction}
J. Song and W. C. Chew, ``Error analysis for the truncation of multipole expansion of vector Green's functions [EM scattering],'' \emph{IEEE Microw. Wireless Compon. Lett.}, vol. 11, no. 7, pp. 311-313, Jul. 2001.
\bibitem{ref_mySYN}
C. Shi, S. Liang, J. Pan, X. Gu, and L. Zuo, ``Generalized Scattering Matrix Synthesis: Independent Region Decomposition for Hybrid
 Antenna--Scatterer Systems,'' \emph{IEEE Trans. Antennas Propag.}, 2025, 10.1109/TAP.2025.3599372.

\bibitem{ref_PRA_truncation_number}
A. Egel, D. Theobald, Y. Donie, U. Lemmer, and G. Gomard, ``Light scattering by oblate particles near planar interfaces: on the validity of the T-matrix approach,'' \emph{Opt. Express}, vol. 24, no. 22, p. 25154, Oct. 2016, doi: 10.1364/OE.24.025154.
\bibitem{ref_Rubio}
J. Rubio and R. Gómez-Alcalá, ``Mutual coupling of antennas with overlapping minimum spheres based on the transformation between spherical and plane vector waves,'' \emph{IEEE Trans. Antennas Propag.}, vol. 69, no. 4, pp. 2103-2111, 2020.


\end{thebibliography}
\end{document}